\documentclass[aps,twocolumn,showpacs]{revtex4}
\usepackage{graphicx}
\usepackage{times,mathptm}

\begin{document}
\title{Air parcel random walk and droplet spectra broadening in clouds}
\author{K.S. Turitsyn}
\email[]{feturits@wicc.weizmann.ac.il}
\affiliation{Physics of
Complex Systems, Weizmann Institute of Science, Rehovot 76100,
Israel}


\begin{abstract}

We study the effect of turbulent flow on the droplet growth in a
cloud during the condensation phase. Using the air parcel model,
we describe analytically how the size distribution of droplets 
evolves at the different stages of parcel movement. We show
that turbulent random walk superimposed on an accelerated ascent
of the parcel makes the relative width of droplet distribution to
grow initially  as $t^{1/2}$ and then decay as $t^{-3/2}$.

\end{abstract}

\maketitle

The growth mechanism of small water droplets in clouds is the 
condensation of water vapour \cite{PK}. While the dynamics of droplet growth
is well understood, no consistent statistical theory exists yet for describing 
the evolution of droplet spectra in the cloud --- see \cite{PK,K} and the 
references therein. The main difficulties in this task are that a consistent theory should
describe the interaction of at least four fields in the cloud:
vapour concentration, temperature, droplet distribution
 and velocity. For sufficiently small droplets, the concentration 
 inhomogeneities due to droplet inertia are
negligible \cite{WM,SE,FFS}. 
In cloud cores, vapour concentration and temperature
also can be considered smooth functions of position, so here we shall
only consider their vertical dependence. It is the velocity field which has
a turbulent nature and can  be considered as a stochastic variable. The 
statistical description of random interacting fields belongs to the class
of field-theoretical problems which are generally unsolvable. In
our case, neglecting the spatial fluctuations of concentrations
and temperature, one can turn the problem into that of a random walk
superimposed on a regular drift in the prescribed environment. Three
limiting cases of that problem (with no drift, permanent-velocity
drift and permanent-acceleration drift) are solved here
analytically. 
This might provide a useful  opportunity for comparison
with the data of observations and numerical simulations. 
We use
the concept of an air parcel \cite{PK}, which means that we consider
a macroscopic volume of air, which is moving through the cloud
with the droplets within it. Together with the deterministic
motion and growth, the parcel participates in a turbulent random
walk which can be modelled as brownian motion on a timescale
larger than the velocity correlation time. We examine how the form
of the droplet size distribution changes with time. It is shown
that the dynamics of the form of the distribution (spectrum)
strongly depends on the deterministic motion. The relative spectral width
decreases while the parcel is moving upwards either with constant
acceleration or with a constant velocity. When there is no vertical drift 
(the parcel 
comes into mechanical equilibrium with its environment
and wanders around a fixed height), the relative width of
the distribution begins to grow. The form of the spectrum increases
in width. We also show, that the distribution of 
droplets over sizes is asymmetric: power law at small sizes part
is determined by the activation of new nuclei while the large sizes part 
is exponential.

\begin{figure*}[tb]
  { \centering
    \includegraphics[width=131pt]{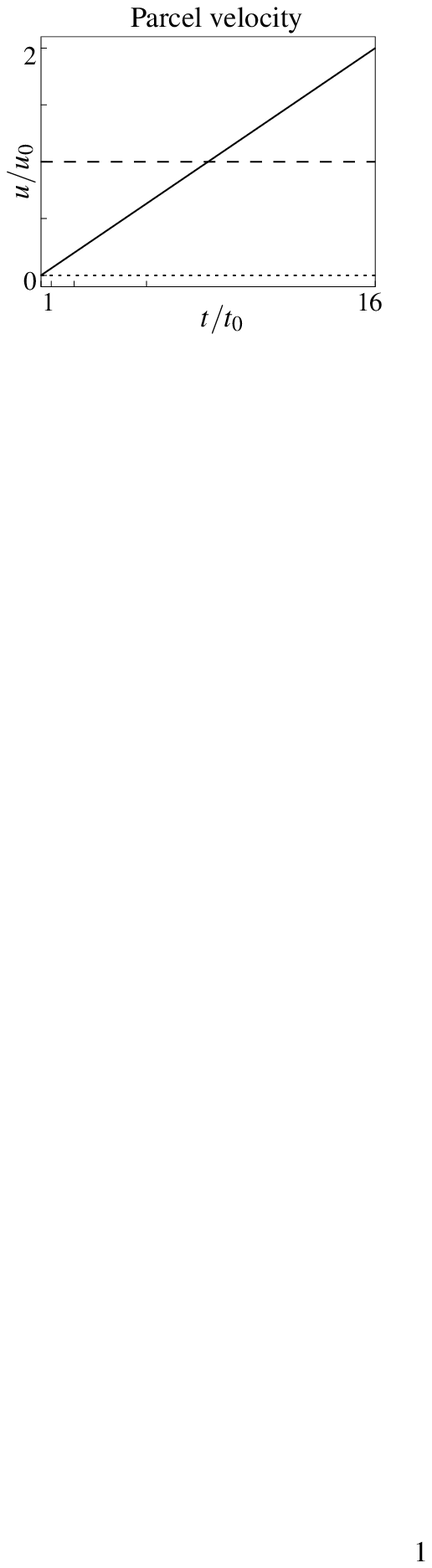}\hspace{30pt}
    \includegraphics[width=131pt]{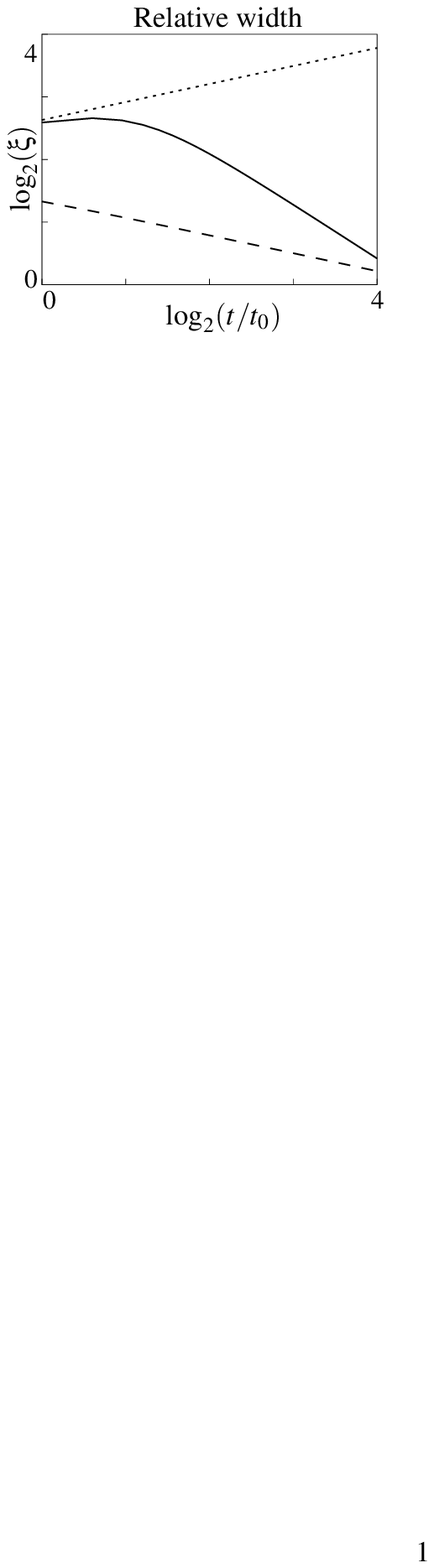}\hspace{30pt}
    \includegraphics[width=131pt]{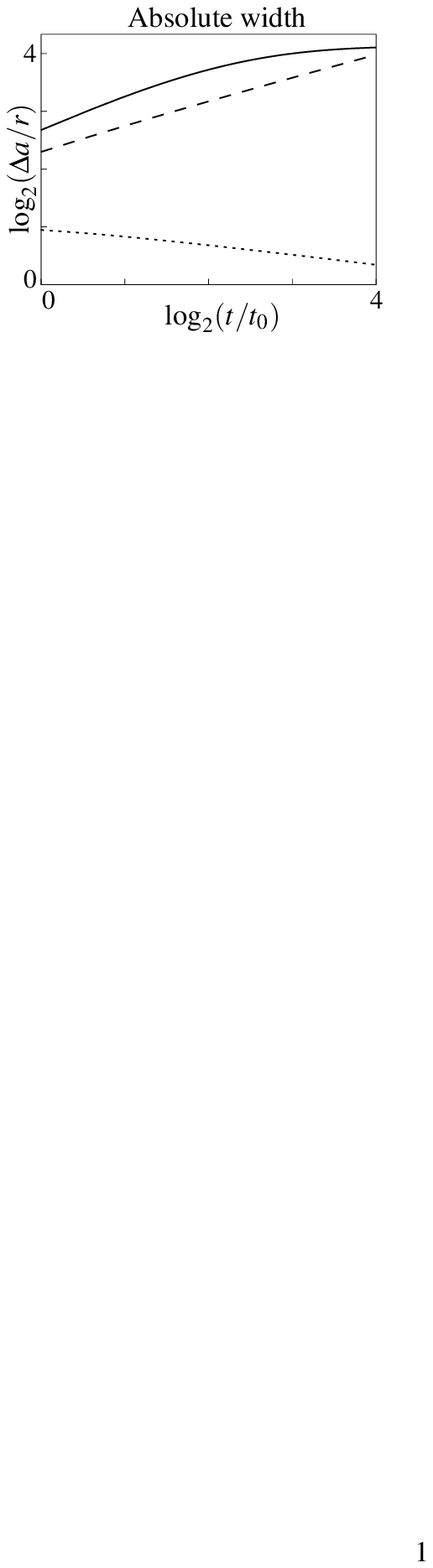}
    \caption{An example of the parcel deterministic velocity
    trajectory together with the relative and absolute
    distribution width for arbitraty chosen parameters $t_0,u_0,r$. 
    The three curves in each plot 
    correspond to the different cases: parcel with constant acceleration
    (solid line), parcel with constant velocity (dashed line) and parcel with 
    zero velocity (dotted line).
    \label{figex}}
  }
\end{figure*}
Advection and 
diffusion of passive fields in turbulent fluid is usually described with
the Kraichnan model (see e.g. \cite{FGW}). According to this model the velocity 
field can be considered a stochastic Markovian process. The stochastic 
description is valid when the fluid motion is chaotic. This is true in the case of 
fully developed turbulence which corresponds to large values of Reynolds 
number \cite{Frisch}. In real clouds the observed Reynolds number is of order 
${\mathrm Re} \sim 10^5\div 10^8$ \cite{PK}, which is certainly enough to assume 
air velocity field stochastic. The validity of the zero correlation time assumption
is proved by the following arguments: correlation times of velocity fluctuations 
do not exceed $10^2$ s, while the duration of the condensational stage lasts
$10^3$ s or more \cite{PK}. Thus a more realistic model, which takes into account the 
finite correlation times would give a small correction not exceeding  $10\%$. In this
paper we focus only on the vertical motion of a particle, because 
horizontal inhomogeneities have fluctuating nature, and will not lead to any
new qualitative effect. We also neglect collisions and coalescence of droplets
by considering the initial stage of growth when droplets are small and 
the only relevant droplet growth mechanism is condensation of 
water vapour\cite{PK}. 
Using all these assumptions, one can write the set
of equations describing the dynamics of a single droplet \cite{PK,K}:
\begin{eqnarray} \label{main}
  && \frac{\mathrm{d} A}{\mathrm{d}\, t}=
    \frac{\mathrm{d}\,a^2}{\mathrm{d}\, t}  = s(z,t), \\
  && \frac{\mathrm{d} z}{\mathrm{d}\, t}  = u(t)+v(t)\ .
    \label{mainwalk}
\end{eqnarray}
Here $a$ is the droplet radius, $A=a^2$ is a variable,
proportional to the droplet surface area, and $s$ is the local
supersaturation rate defined as $s=n/n_0-1$, where $n$ and $n_0$
are the local vapour concentration and the concentration of saturated vapour 
correspondingly. We have set the coefficient of propotionality
to $1$, which can be always achieved by using the proper dimensional units. 
For any given droplet, the supersaturation rate changes
with time because the air parcel containing this droplet is
lifted up to the cloud and is being cooled along the way. The
vertical velocity of a parcel (\ref{mainwalk}) consists of two
components: the deterministic $u(t)$, which corresponds to the
macroscopic updraft flow and the random $v(t)$ which is due
to turbulent velocity fluctuations. We will consider the random
velocity to be gaussian and short correlated
\begin{equation}\label{stat}
 \left\langle v(t)v(t')\right\rangle = D \delta(t-t').
\end{equation}
In the following we will analyze the dynamics of the droplet distribution
$P(A,z,t)$. The quantity $P(A,z,t)\mathrm{d} A\mathrm{d}z$ represents the
fraction of the droplets which are located in the area $(z,z+\mathrm{d}z)$ and have
the surface area in the interval $(A,A+\mathrm{d}A)$. As long as we do
not take into account the interaction of the droplets, the dynamical equation
governing the evolution of the spectrum is described by the Liouville equation
that corresponds to the single droplet dynamics (\ref{main},\ref{mainwalk}):
\begin{equation}\label{liouville}
 \left\{\partial_t + s(z,t)
 \partial_A + \left[u(t)+v(t)\right] \partial_z \right\} P(A,z,t) =0.
\end{equation}
Applying the standard procedure of averaging over the white noise (\ref{stat})
one obtains the Fokker-Planck equation: 
\begin{equation}\label{fokker}
 \left\{\partial_t + s(z,t)
 \partial_A + u(t) \partial_z - {\textstyle\frac{1}{2}} D
 \partial_z^2\right\}P(A,z,t) =0.
\end{equation}
The second and third terms describe the advection of the droplets in
phase space due to deterministic terms in equations 
(\ref{main},\ref{mainwalk}). They are responsible for the motion of
the distribution center. The third term comes from the stochastic
part of the velocity and its main effect is the spectrum broadening.
We assume that the initial droplet size is small so that we can set
$P(A,z,0) = \delta(A)\delta(z)$. This equation can not be solved
for general supersaturation field $s(z,t)$. However it has an
analytic solution for a stationary linear profile $s(z,t)=s_0+s_1
z$ which we consider as a reasonable local approximation of the real supersaturation
field. The expression for $P(A,z,t)$ is rather bulky. However 
the size distribution $P(A,t)=\int \mathrm{d}z \, P(A,z,t)$ 
is described by the simple expression
\begin{equation} \label{sol}
 P(A,t) = \frac{1}{\sqrt{2\pi \,\delta A^2(t)}}
 \exp\left[-\frac{(A-A_0(t))^2}{2\,\delta A^2(t)}\right].
\end{equation}
Here $A_0(t) = \int_0^t \mathrm{d}t'\left(s_0+ s_1
z_p(t') \right)$ and $z_p(t)$ is a a deterministic parcel
trajectory: $z_p(t)=\int_0^t \mathrm{d}t' u(t')$. The distribution
width is $\delta A^2 (t) = D\, s_1^2 t^3/3$. Usually one is
interested in the distribution of radius $a$; from
(\ref{sol}) we see that it will be substantially non-gaussian in our
case as $P_a(a,t) = 2 a P(a^2,t)$. Nevertheless, it will still have
the bell-form and can be described by it's center $a_0(t) = \sqrt{A_0}$
and width 
\begin{eqnarray} 
 \label{width}
 \delta a(t) = \sqrt{A_0 + \delta A} - \sqrt{A_0} =
 \frac{\delta A}{\sqrt{A_0 + \delta A} + \sqrt{A_0}}.
\end{eqnarray}

One can see that in some situations the distribution in $a$
can be narrowed, even if it is broadened for the variable
$A$. The growth rate of the distribution width can be described by
the variable $\xi = \delta A/A_0$. The relative width of the 
distribution over radii is then
\begin{equation} \label{relation}
 \frac{\delta a}{a_0} = \frac{\xi}{\sqrt{1+\xi}+1}.
\end{equation}
The r.h.s. of this relation is a monotonic function, so the
relative width of the distribution over $a$ behaves qualitatively in the
same way as the variable $\xi$. Even the regular (non-turbulent)
part of the parcel motion in a real cloud can
be very complicated. At the beginning the parcel is being lifted
up the cloud because of the convection mechanism. The parcel is
accelerated until the velocity reaches some stationary value due
to the drag force. During the motion the temperature of the parcel
changes because of droplet growth and heat exchange. When the
parcel temperature becomes equal to the ambient temperature, its
velocity decreases to zero and it stays at this steady height.
One can analyze such a movement by considering three limiting
cases: when the parcel is accelerated so $u(t)=u_0 + \textbf{a}
t$; when the parcel's velocity is constant: $u(t)=u$; and when the
parcel has reached its steady height, and its velocity is zero,
$u(t)=0$. In the following sections we will analyze these three 
cases and show that they correspond to the different types of behaviour
of the size distribution function.

We start
from the simplest case of zero velocity. As already
mentioned, this case is important when analyzing the last
part of the parcel's movement. On this part the distribution
center has already moved from $A=0$ and its width is also finite.
However if the steady stage lasts for a long time, the initial
conditions are forgotten, and one can treat this problem with the
initial conditions $A(0)=0$ and $\delta A(0)=0$. The solution of
this problem was introduced in the previous section (\ref{sol}),
and one obtains: $z_p(t)=0$, $A_0(t)=s_0 t$,
$\delta A^2(t) =D s_1^2 t^3/3$. Thus for the relative
width one finds:
\begin{equation} \label{rel0} \xi =
 \frac{\delta A}{A} = \sqrt{D/3}\,\frac{s_1}{s_0}\, t^{1/2}.
\end{equation}
Therefore the distribution is being broadened  with
a rate $\xi \propto t^{1/2}$ at this stage. Turning to the
radius distribution, $P_a(a,t)$, one can say that its relative
width is increasing with the same rate $\delta a/a \propto \xi
\propto t^{1/2}$ for small times $t < s_0^2/s_1^2 D$, but as $\xi$
becomes large enough according to (\ref{relation}) its growth
rate changes to $\delta a/a \propto \xi^{1/2} \propto t^{1/4}$.
The radius distribution is broadened slower than that of the surface area.
It is worth noting that in stratiform clouds where the
deterministic parcel movement is small the zero velocity case is
most important. We see,that one should observe strongly broadened
distributions there.

Next we consider the constant velocity case: $z_p(t)= u t $,
$A_0(t) = A_0(0)+s_0 t + (s_1 u /2)t^2$. Again, at
large enough times, the distribution forgets about the initial
conditions (which were formed during the acceleration stage) and the 
last term in $A_0(t)$ will become dominant, so $A_0 = (s_1 u /2)t^2$,
$\delta A^2 = (D s_1^2/3) t^3 $. Therefore the relative
distribution width is slowly decreasing: $\xi \propto t^{-1/2}$,
while the center of the distribution and its absolute width are
growing with the rates $a_0 = \sqrt{A_0} \sim t$, $\delta a \sim
\sqrt{A_0}\,\xi \propto t^{1/2}$.

Finally, we consider a uniformly accelerating parcel. Here one finds
$s(t)=s(z_p(t)) = s_0 + s_1 u_0 t + s_1 \textbf{a} t^2$. If
initial droplets sizes are small, we can easily find the dynamics
of the center of drops surface area distribution, $A_0(t) = s_0 t + (s_1
u_0/2)t^2+ (s_1 \textbf{a} /6) t^3$. The relative width behaves
like
\begin{equation} \label{rel1}
 \xi = \frac{\delta A}{A_0} =
 \frac{\sqrt{D s_1^2/3}\, t^{3/2}}{s_0 t + (s_1 u_0/2)t^2+ (s_1
 \textbf{a} /6) t^3}.
\end{equation}
At the very small times the relative width is growing with the
rate $\xi \propto t^{1/2}$, as in the zero velocity case. At large
times, when the parcel has accelerated enough, the relative width
tends to zero as $\xi \propto t^{-3/2}$. Therefore, there is
a time when the relative width has a maximum. The position of the
maximum is determined by the values of $u_0,\textbf{a}$. It can be
estimated as $t_{\rm max}\sim \min\left[s_0/(s_1
u_0),\sqrt{s_0/(s_1 \textbf{a})}\right]$. The value of the maximum
is in this case $\xi_{\rm max} \sim \sqrt{D t_{\rm max}}s_1/s_0$.
It is interesting, that the absolute width $\delta a =
\sqrt{A_0}(\sqrt{1+\xi}-1) \sim \sqrt{A_0}\,\xi$ is always
increasing, and approaches a constant asymptotic value at large
times. The particular value of this width is a complicated
function of all the parameters.

The figure \ref{figex} shows the qualitative behaviour of the absolute and
relative width for the three analyzed cases. The pictures should not be 
considered as a prediction for some particular cloud, but as an example,
which shows how different velocity trajectory can affect the distribution
parameters.

So far we have
analyzed the situation where small droplets are injected into
the cloud only once, at the beginning. In real clouds the situation is
certainly different. New nuclei are activated all the time, and
their activation surely affects the distribution form.
Nevertheless, we will show that this effect of new nuclei
activation does not change the large droplet sizes part of the distribution. 
The droplet size distribution can be written as follows:
\begin{equation} \label{real}
 n(a,t) = \int_0^t \mathrm{d} t' n_0(t') P_a(a,t-t')\,,
\end{equation}
where $n_0(t)\mathrm{d} t$ is
the number of nuclei activated in the interval $(t,t+\mathrm{d}
t)$. Luckily the theory of nuclei activation is well developed 
\citep{PK}. The simplest model of nuclei activation predicts, that
the activation rate is proportional to some power of local
supersaturation rate, $n_0(t) \propto s^{-k}(t)$, and usually
$k>0$. The absolute width $\delta a(t)$ of the distribution
$P_a(a,t)$ is a growing function, therefore the front of the real
distribution will be always determined by the droplets which began
to grow at $t=0$. Nevertheless the new droplets determine the form
of the left part. Assuming the distribution to be narrow enough,
one can have the following approximation to the small droplet part 
of the spectrum:
$n(a_0-\Delta a,t) \sim n_0(t^*)$, where $t^*$ is determined by
the condition that $P_a(a,t^*)$ should be centered at $a-\Delta
a$. For example for the acceleration stage $s(t) = s_0 + s_1
\textbf{a} t^2/2 $, and the center position is given by $A_0(t) =
s_0 t + s_1 \textbf{a} t^3/6 \propto  t^3 $. Thus $a_0(t) \propto
t^{3/2}$. From the definition of $t^*$ we find $t-t^*\propto
(a_0-\Delta a)^{2/3}$, so $t^* \propto a_0^{2/3}-(a_0-\Delta
a)^{2/3}\propto a_0^{-1/3} \Delta a$ for small $\Delta a \ll a_0$.
Recalling that $n_0(t) \propto s^{-k}$, we find that $n(a_0-\Delta
a,t) \propto \left(s_0 +c\, \Delta a^2\right)^{-k}$. We can see
that the left part will have a rather wide power law tail
distinct from a narrow exponential right tail. This
is also an important result, which might be observed in real clouds.

To conclude, we have shown that the evolution of 
size distribution of droplet spectra
can be divided into three different stages. At each stage,
the relative width of the  distribution  behaves according to different
power  laws: it grows with a rate $\xi \propto t^{1/2}$ at the
zero velocity  stage, and decays with the laws $\xi \propto t^{-1/2},
\xi \propto t^{-3/2}$ on the stages with parcel constant velocity,
constant acceleration respectively. The dynamics of real air parcels
is  more complicated, and the equation (\ref{sol})  allows one to estimate
the distribution width according to our model. However, from
physical reasoning we see that real parcel dynamics can be
approximated by three consequent stages, where the relative
width asymptotics were computed analytically.

We have also shown, that the left part of droplet size spectra is mainly
determined by the process of nuclei activation during the initial
stage of parcel movement. In this case one will observe a power-law
tail in the small droplets part of spectrum, in contrary to 
exponential tail for large droplet sizes, which is determined by the parcel's 
velocity fluctuations.

{\em Acknowledgements:} Author would like to thank G.~Fal\-kovich
for the most fruitful and inspiring discussions and M.G.~Stepanov
for useful remarks and technical assistance in making figures.

 \end{document}